**A two-terminal spin valve device controlled by spin-orbit torques with enhanced giant magnetoresistance**

Can Onur Avci[*], Charles-Henri Lambert, Giacomo Sala, Pietro Gambardella

*Department of Materials, ETH Zürich, CH-8093 Zürich, Switzerland*

[*]corresponding author: can.onur.avci@mat.ethz.ch

**We report on the combination of current-induced spin-orbit torques and giant magnetoresistance in a single device to achieve all-electrical write and read out of the magnetization. The device consists of perpendicularly magnetized TbCo and Co layers separated by a Pt or Cu spacer. Current injection through such layers exerts spin-orbit torques and switches the magnetization of the Co layer while the TbCo magnetization remains fixed. Subsequent current injection of lower amplitude senses the relative orientation of the magnetization of the Co and TbCo layers, which results in two distinct resistance levels for parallel and antiparallel alignment due to the current-in-plane giant magnetoresistance effect. We further show that the giant magnetoresistance of devices including a single TbCo/spacer/Co trilayer can be improved from 0.02% to 6% by using a Cu spacer instead of Pt. This type of devices offers an alternative route to a two terminal spintronic memory that can be fabricated with moderate effort.**

Spin-orbit torques (SOTs) have emerged as a versatile tool to manipulate the magnetization in magnetic heterostructures[1-10]. They typically occur in normal metal/ferromagnet (NM/FM) bilayers where the bulk spin Hall and interfacial spin galvanic effects convert the injected charge current into pure spin currents[11]. Because there is no need for a FM polarizer to generate spin currents, in contrast to the conventional spin-transfer torque schemes[12], SOTs offer great flexibility in device design and functionality[13,14]. One such device is the three-terminal magnetic tunnel junction (MTJ), where the magnetization of the free FM layer is controlled by planar current injection generating SOTs, and the magnetization state is probed by vertical current injection through the oxide barrier via the tunnel magnetoresistance[5,15,16]. Three-terminal MTJs are considered for scalable, low-power and high speed magnetic random-access memory (MRAM) applications[17-21]. Additionally, SOTs allow for the realization of even simpler two terminal memory devices[22,23]. These could complement the MRAM production process and diversify the circuit design, material spectrum, and related physical phenomena.

In this work, we describe a two-terminal device where the magnetic state is controlled and probed by currents sent through the same planar path. This simple device is made of a hard FM (TbCo) and a soft FM (Co), each possessing perpendicular magnetic anisotropy, separated by a nonmagnetic spacer layer. The Co layer is in contact with Pt, which acts as an SOT generator. Current injection in the presence of a static in-plane field switches the magnetization of Co between up and down states as in the standard SOT switching scheme[1]. However, unlike in the typical SOT devices, the magnetization state is not probed by the anomalous Hall effect but rather by the current-in-plane giant magnetoresistance[24,25], which has two distinct levels for parallel and antiparallel orientations of TbCo and Co. In the following, we provide a proof of concept demonstration of such a device and show how to improve the magnetoresistance ratio in planar SOT structures including Pt. Our concept offers an alternative path to a highly scalable, all-electrical, two terminal memory that can be produced with minimum fabrication efforts.

We deposited //Ti(3)/TbCo(16)/Co($t_{Co}$)/Pt($t_{Pt}$)/Co(1.4)/Ti(5) and //Ta(2)/Pt(4)/Co(1.2)/Cu(3.1)/Co(1.2)/TbCo(16)/Ti(5) by d.c. magnetron sputtering onto a Si/SiO$_2$ substrate at room temperature. The numbers in brackets correspond to the thickness in nm and the composition of TbCo was 27% Tb and 73% Co. The bottom Ti or Ta and the top Ti serve as buffer and capping, respectively. In the first batch of samples, the thin Co layer in direct contact with TbCo is used to enhance the magnetoresistance effect, as explained later, and will be considered as part of the TbCo layer and not mentioned explicitly (unless stated otherwise). The thickness of this coupling layer was $t_{Co} = 0$, 0.4 and 1.2 nm and the thickness of the Pt spacer was varied between $t_{Pt} = 3$ nm and 4.5 nm in steps of 0.25 nm for $t_{Co} = 0.4$ nm. All layers were grown in a base pressure of $\sim 5 \times 10^{-8}$ mbar and Ar partial pressure of $2 \times 10^{-3}$ mbar. Two-terminal 2.5-μm-wide 7.5-μm-long current tracks as well as 5 μm-wide 20-μm-long Hall bars were defined on blanket substrates using UV photolithography. After deposition, the devices were obtained by lift-off as depicted in Fig. 1 (a). For electrical characterization of the magnetization and for probing the magnetization state during the switching experiments, we measured the Hall effect and longitudinal resistance using standard a.c. current injection methods[26]. All experiments were performed in ambient conditions.

We first focus on the first batch of samples with the Pt(4.5 nm) spacer. Figure 1 (b) shows the Hall resistance ($R_H$) of the sample with the Pt(4.5 nm) spacer during an out-of-plane field ($B_z$) sweep. As evident from the data, two separate magnetization reversal events occur at low field (~25 mT) and at high field (~750 mT). The low field reversal is attributed to the top Co layer as it is within the typical coercivity range (10-50 mT) expected of single Pt/Co bilayers. The high field reversal is attributed to TbCo, which is known to possess large perpendicular magnetic anisotropy and large coercivity near the magnetic compensation point. This measurement shows that the two layers are magnetically decoupled and that their relative magnetic orientation can be preset by an external field. Another interesting aspect is the sign of the reversal events. Sweeping from $+B_z$ to $-B_z$ (blue arrows), $R_H$ first decreases at small negative fields, corresponding to Co switching from up to down, then increases when TbCo switches from up to down. The increase of $R_H$ when TbCo switches from up to down indicates that the magnetization is dominated by the Tb sublattice, since $R_H$ is mainly due to the Co sublattice[27]. The reversed sweep signal (red arrows) can be explained using the same arguments.

We then measure the longitudinal resistance ($R$) during a sweep of $B_z$ [Fig. 1 (c)]. For these measurements we used the two-terminal device shown in the bottom panel of Fig. 1 (a) –, fabricated on a separate chip. Interestingly, we observe a spin-valve-like signal that depends on the relative magnetic orientation of the layers, i.e., two reversal events at low and high fields, which give rise to two distinct resistance levels. We interpret this behavior as the current-in-plane giant magnetoresistance (GMR) mediated by the Pt spacer. The GMR, however, is about 0.03%, two orders of magnitude smaller than typical values for TbCo-based multilayers using a Cu spacer[28,29]. This low value of the GMR is attributed to the strong spin scatter properties of Pt[30,31]. Unlike in the conventional GMR measurements, the resistance is high when Co and TbCo are parallel to each other and low when they are antiparallel. The high resistance for the parallel state indicates that the GMR mainly originates from the interplay between the magnetization of the top Co layer and the Co magnetic sublattice in the TbCo layer, in agreement with previous studies of the GMR in rare-earth transition-metal alloys[32-34]. Because the magnetization in TbCo is Tb-like, in the parallel (high resistance) configuration the magnetization of the top Co layer is antiparallel to the Co sublattice in the bottom layer, which gives rise to the larger resistance. The dominant GMR contribution of the Co sublattice in TbCo is reasonable because the spin-polarized conduction is mostly dominated by the $s$-$d$ electrons of Co, whereas the spin-polarized $f$ electrons of Tb have negligible contribution to the conduction[35]. The same argument also explains the negative $R_H$ behavior discussed above.

Figure 1 (d) and (e) show the minor loops corresponding to the reversal of the Co magnetization detected by the Hall effect and resistance, respectively, for two different orientations of TbCo magnetization ($\mathbf{m}_{TbCo}$). In the $R_H$ data [Fig. 1 (d)], the loops are nearly identical for both positive and negative $\mathbf{m}_{TbCo}$ with an offset originating from the $R_H$ of TbCo. In the resistance data [Fig. 1 (e)] however, the loops have opposite sign for positive and negative $\mathbf{m}_{TbCo}$ as the orientation of $\mathbf{m}_{TbCo}$ acts as a reference for the resistance readout. This sign reversal will be important to understand the SOT-induced switching data discussed below.

Next, we perform current-induced switching experiments on the resistance tracks. Figure 2 (a) shows the resistance during a pulsed-current sweep at $j < |3.2 \times 10^7|$ A/cm$^2$ (estimated by considering the full stack but excluding the Ti buffer and cap) in the presence of an in-plane field $B_x \sim 200$ mT applied along the current direction as required for SOT switching[1]. Note that for each data point in this and the following switching measurements we have injected a train of five 50-ns-long pulses and averaged the resistance signal for 1 second. This pulsing/averaging scheme minimizes the signal drift. We observe that the magnetization of Co switches between the up and down states reversibly at around 2.5-3x10$^7$ A/cm$^2$. In reference to Fig. 1 (e), for the same (opposite) field and current direction, the Co magnetization switches from up to down (down to up), which is consistent with the switching of Pt/Co bilayers dominated by the damping-like SOT[1,36]. We note that the changes in the resistance upon the SOT-induced switching is about half of that reported by field-induced switching in Fig. 1 (e). This occurs because only the narrow part of the track, where the current density is maximum, switches due to the SOT, whereas the field can also switch the wider contact regions, which also contributes to the resistance.

To test the reliability of the device over multiple switching events, we injected consecutive pulse trains of opposite polarity with $B_x = 200$ mT. Figure 2 (b) shows the pulse sequence (upper panel) and the resistance readout after each pulse (lower panel). We observe that the switching is fully reproducible for over 40 switching events and no drift or partial switching is observed. Moreover, we tested the $B_x$ dependence of $j_{crit}$ in a wide range of fields and confirmed the behavior expected of SOT-switching. The field dependence of the critical switching current threshold ($j_{crit}$) is plotted in Fig. 3. We observe that $j_{crit}$ decreases monotonously with increasing $B_x$, as the energy barrier between the up and down states is effectively reduced. $j_{crit}$ increases more significantly when $B_x < 100$ mT, which is also expected because $j_{crit}$ is found to diverge as $B_x \to 0$ (Ref. [36]). We were unable to extend the measurements to $B_x < 50$ mT because Joule heating due to the increased $j$ was negatively influencing the readout, and also, partial switching of TbCo was observed in some cases.

Starting from our proof of concept for a two-terminal spin valve device operated by SOT, we discuss several pathways to enhance the signal output. One way to enhance the readout signal is to increase the GMR through interface engineering. In Fig. 4 (a) we compare the relative resistance change ($\Delta R/R$) in three different samples with and without a Co insertion layer in between TbCo and Pt. We observe that $\Delta R/R$ increases from 0.02 to 0.05% as the Co thickness increases from 0 to 1.2 nm. Moreover, we notice that the coercivity of the reference layer is significantly larger when Co is introduced between TbCo and Pt, which is indicative of an enhanced PMA. This comparison shows that for optimum readout, an insertion layer is beneficial to enhance both the interface contribution to the GMR and the PMA of the reference layer. Further enhancement of the GMR is possible, in principle, by modifying the interface between the top Co layer and Pt. However, this approach would affect also the SOT acting on the free layer, because the SOT properties depend sensitively on this interface[37].

Another path to increase the readout signal is to reduce $t_{Pt}$ to maximize the GMR effect[38]. However, for the symmetric SOT switching operation (i.e., the same $j_{crit}$ switches from the parallel to the antiparallel state and vice versa) the two layers need to be magnetically decoupled. Figure 4 (b) shows the Co minor loop data for positive $\mathbf{m}_{TbCo}$. We observe that the Co hysteresis loops are significantly influenced by $t_{Pt}$. For $t_{Pt}$ between 3.0 and 3.5 nm, the hysteresis occurs only for positive $B_z$, meaning that the coupling is antiparallel and the effective coupling field is larger than the coercivity of the top Co layer. The magnitude of the coupling gradually decreases with increasing $t_{Pt}$ and the hysteresis loop becomes fully symmetric with respect to $B_z = 0$ mT at $t_{Pt} = 4.5$ nm. Figure 4 (c) summarizes our findings by showing the coercivity (left axis) and the coupling strength (right axis) characterized by the shift of the bistable region delimited by the reversal events as a function of $t_{Pt}$. We observe that while the coercivity is fairly constant, the coupling strength dramatically changes as a function of $t_{Pt}$ within the studied range. We delimit three regions: *i)* strong coupling where the interlayer coupling is larger than the coercivity of the minor loop, hence no bistable region exists at $B_z = 0$ mT (not suitable for SOT switching); *ii)* weak coupling where the interlayer coupling is lower than the coercivity but large enough to prevent SOT from switching; *iii)* negligible coupling where the interlayer coupling is barely noticeable hence the minor loop is symmetric around $B_z = 0$ mT. Based on these measurements we conclude that we can only use $t_{Pt} \geq 4.25$ nm, which explains the small GMR of this type of devices. We speculate that replacing the Pt spacer by a finely tuned light metal/Pt bilayer, where the light metal (e.g., Cr, Ti or Cu) is inserted between TbCo and Pt, could generate a large enough SOT to switch the top Co layer and increase the GMR readout signal significantly[39,40].

Finally, we show that by using a Cu spacer and placing the Pt layer on the opposite side of the Co interface relative to TbCo we obtain a strong increase of the GMR while preserving the SOT switching capability. Figure 5 (a) shows the magnetoresistance of Pt(4)/Co(1.2)/Cu(3.1)/Co(1.2)/TbCo(16) during a $B_z$ field sweep. Similar to the data in Fig. 1 (c), we observe a spin valve behavior but with a much higher GMR ratio reaching up to 5.9%, which is about two orders of magnitude larger compared to the samples with the Pt spacer. The enhanced GMR is a consequence of the use of a relatively thin Cu spacer instead of a thick Pt spacer and the relatively thick Co(1.2 nm) coupling layer attached to TbCo, in view of the conclusions drawn from Fig. 4 (a). Figure 5 (b) reports the current-induced switching measurements on this sample. We achieve switching with moderate currents of about $j \sim 1$-$1.5 \times 10^7$

A/cm$^2$. We observe that for both orientations of $\mathbf{m}_{TbCo}$ the switching is partial (~70%) compared to the field-driven switching shown in Fig. 5 (a). We assume that the partial switching is related to the inhomogeneous currents flowing in different regions of the Hall bar. These data unambiguously demonstrate a two-terminal spin valve device that is controlled by SOTs with high read out GMR and moderate write current requirements.

The magnetoresistive reading of spin valves with in-plane or out-of-plane magnetization is a paradigmatic feature of spintronic devices[41,42]. Usually, the magnetization in such structures is controlled by either an external field or the spin transfer torques due to current injection perpendicular to the magnetic layers[43,44]. Another application of such devices is to study domain wall propagation in the soft layer either by fields or spin-transfer torques while the hard layer remains in a single-domain state serving as reference[45-47]. Only recently there have been efforts to include SOT switching in spin valve devices[48,49]. However, these studies employed magnetically coupled layers and the Hall effect for reading their magnetic state, which is not suitable for two-terminal devices. Our measurements take full advantage of the spin valve functionalities, adding the versatility of SOT switching to the simplest magnetoresistive reading of the magnetization . Furthermore, our concept can enhance the functionality of earlier domain wall devices by taking advantage of strong SOTs for efficiently nucleating and driving domain walls[7,8]. One use of such device could be to generate analogue-like resistive signal outputs that depend on the domain wall position. This device concept could find interest in neuromorphic computing applications[50].

**Data Availability**

The data that support the findings of this study are available from the corresponding author upon reasonable request.

**Acknowledgments**

This work was supported by the Swiss National Science Foundation through grants #200020_172775 and PZ00P2-179944.

**Figures**

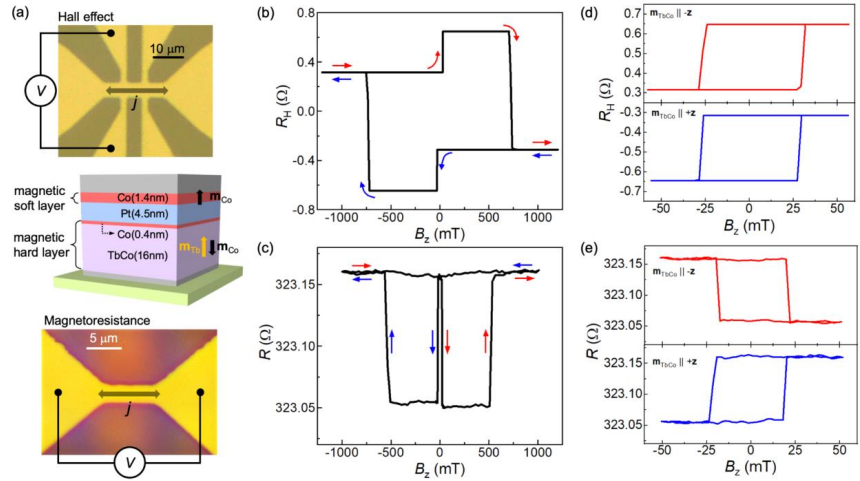

**Figure 1** - (a) Schematics of the devices and multilayer structure used in this study. (b) Hall resistance ($R_H$) and (c) magnetoresistance ($R$) response of the devices shown in (a) during an out-of-plane field ($B_z$) sweep. (d) Minor loops corresponding to the magnetization reversal of the top Co layer for the two different orientations of TbCo. The thickness of the Pt spacer in (b-e) is 4.5 nm. See text for further details.

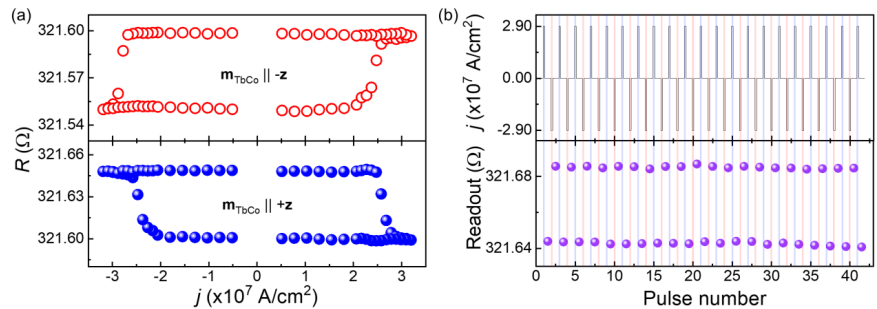

**Figure 2** - (a) Longitudinal resistance (R) measurement of current-induced magnetization reversal of the top Co layer for negative (top panel) and positive (lower panel) magnetization of TbCo. We observe reversible switching events starting at around 2-5 – 3 x $10^7$ A/cm$^2$ for both TbCo magnetizations and positive/negative current polarity. (b) Consecutive switching of the magnetization over 40 pulse cycles. The in-plane field was set to +200 mT in both sets of measurements. For each data point a train of five pulses of 50 ns duration was sent through the device. The data in each loop were averaged over 10 full cycles to enhance the signal to noise ratio.

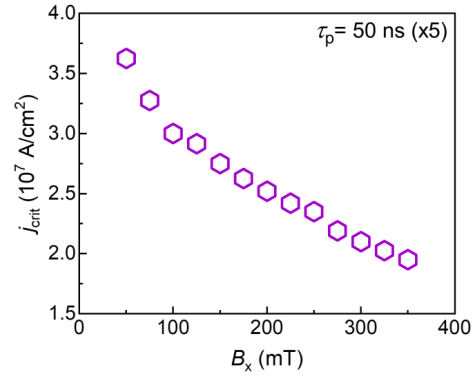

**Figure 3** - Dependence of critical switching current density ($j_{crit}$) as a function of the static in-plane field ($B_x$).

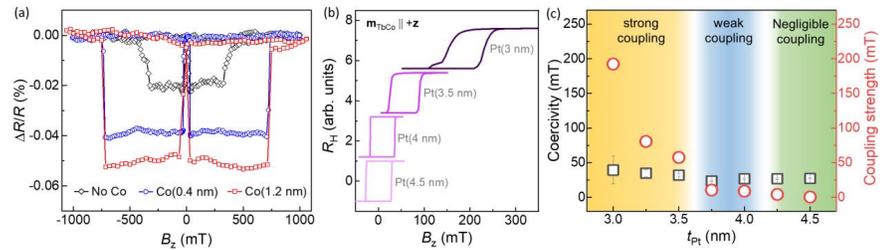

**Figure 4** - (a) Effect of Co insertion layer on the magnetoresistance ($\Delta R/R$). (b) Co minor loops measured using the Hall resistance ($R_H$) for different thickness of the Pt spacer ($t_{Pt}$). (c) Coercivity of the Co layer (squares, left axis) and coupling strength (circles, right axis) as a function of $t_{Pt}$. We identify three regions with strong, weak and negligible coupling. See text for more details.

Applied Physics LettersACCEPTED MANUSCRIPT

This is the author's peer reviewed, accepted manuscript. However, the online version of record will be different from this version once it has been copyedited and typeset.

PLEASE CITE THIS ARTICLE AS DOI: 10.1063/5.0055177
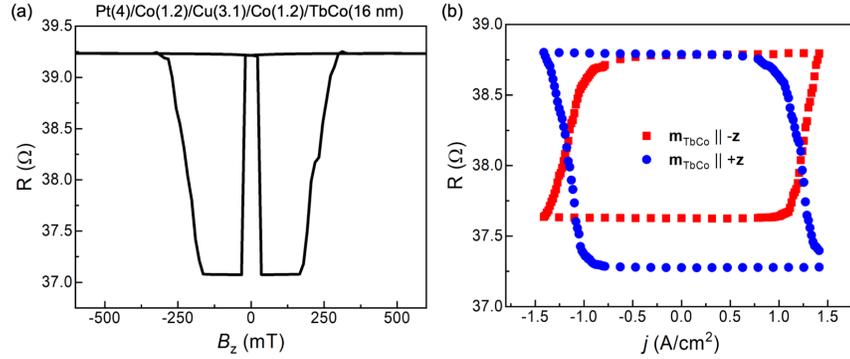

**Figure 5** – (a) Magnetoresistance as a function of out-of-plane magnetic field of a spin valve consisting of Pt(4)/Co(1.2)/Cu(3.1)/Co(1.2)/TbCo(16) with a Cu spacer instead of Pt. (b) Current-induced switching of the bottom Co layer induced by the SOT generated from the Pt underlayer. The magnetoresistance is measured by a four-point probe on a Hall bar structure similar to the one shown in the upper panel of Fig.1 (a). For the switching measurements we have used pulses of 10 ms and averaged the data over three consecutive cycles. The in-plane field $B_x$ was set to +300 mT.

## References

[1] Ioan Mihai Miron, Kevin Garello, Gilles Gaudin, Pierre-Jean Zermatten, Marius V. Costache, Stéphane Auffret, Sébastien Bandiera, Bernard Rodmacq, Alain Schuhl, and Pietro Gambardella, Nature **476** (7359), 189 (2011).

[2] Can Onur Avci, Kevin Garello, Corneliu Nistor, Sylvie Godey, Belén Ballesteros, Aitor Mugarza, Alessandro Barla, Manuel Valvidares, Eric Pellegrin, Abhijit Ghosh, Ioan Mihai Miron, Olivier Boulle, Stephane Auffret, Gilles Gaudin, and Pietro Gambardella, Physical Review B **89** (21) (2014).

[3] K. Garello, I. M. Miron, C. O. Avci, F. Freimuth, Y. Mokrousov, S. Blugel, S. Auffret, O. Boulle, G. Gaudin, and P. Gambardella, Nat Nanotechnol **8** (8), 587 (2013).

[4] J. Kim, J. Sinha, M. Hayashi, M. Yamanouchi, S. Fukami, T. Suzuki, S. Mitani, and H. Ohno, Nat Mater **12** (3), 240 (2013).

[5] Luqiao Liu, Chi-Feng Pai, Y. Li, H. W. Tseng, D. C. Ralph, and R. A. Buhrman, Science **336** (6081), 555 (2012).

[6] Chi-Feng Pai, Luqiao Liu, Y. Li, H. W. Tseng, D. C. Ralph, and R. A. Buhrman, Applied Physics Letters **101** (12), 122404 (2012).10AIP Publishing

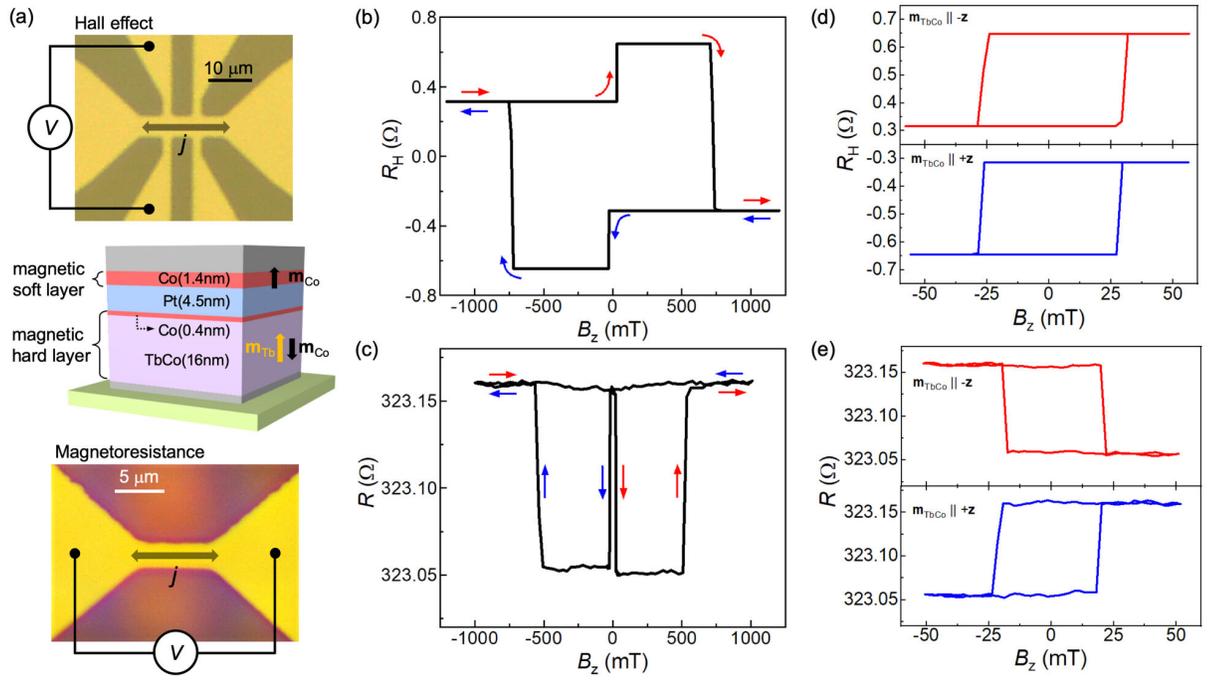

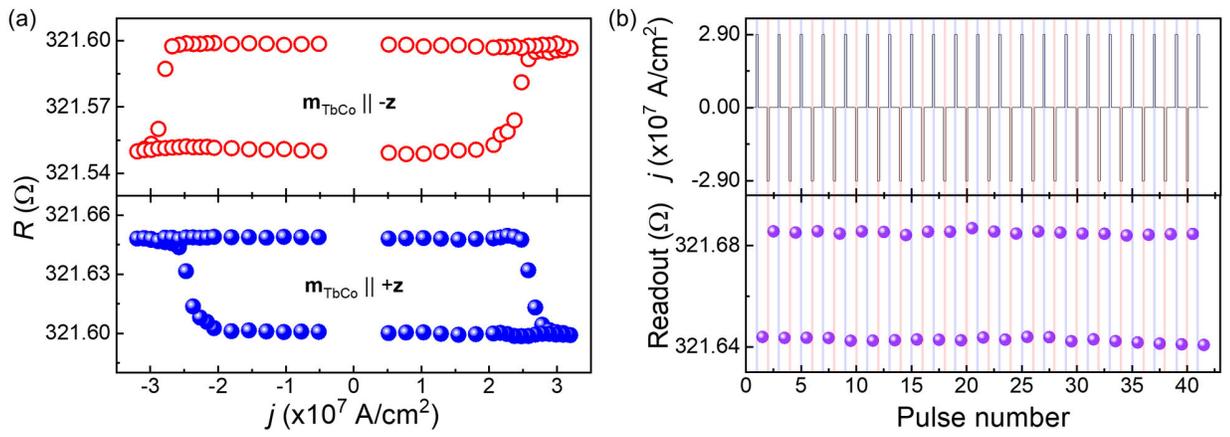

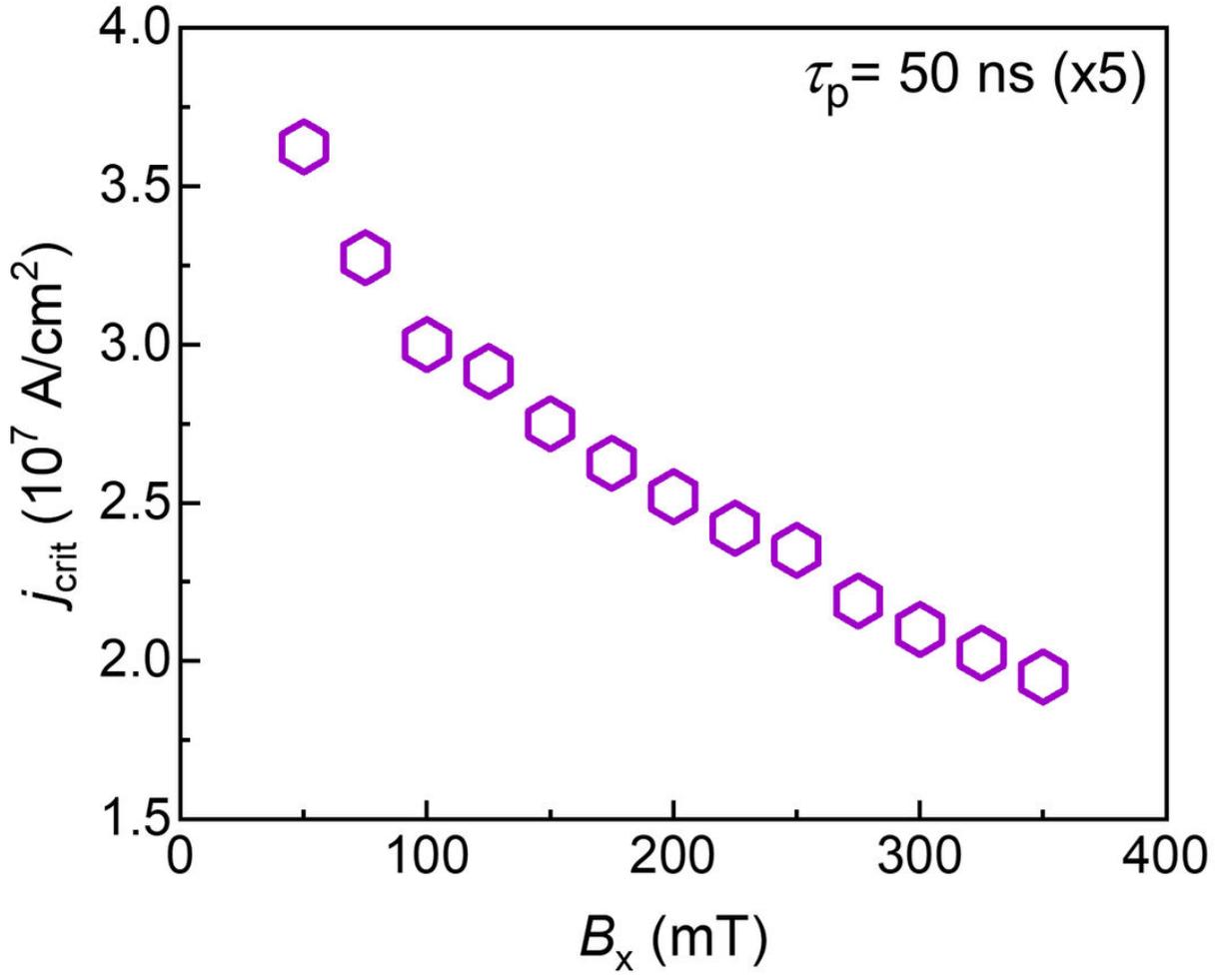

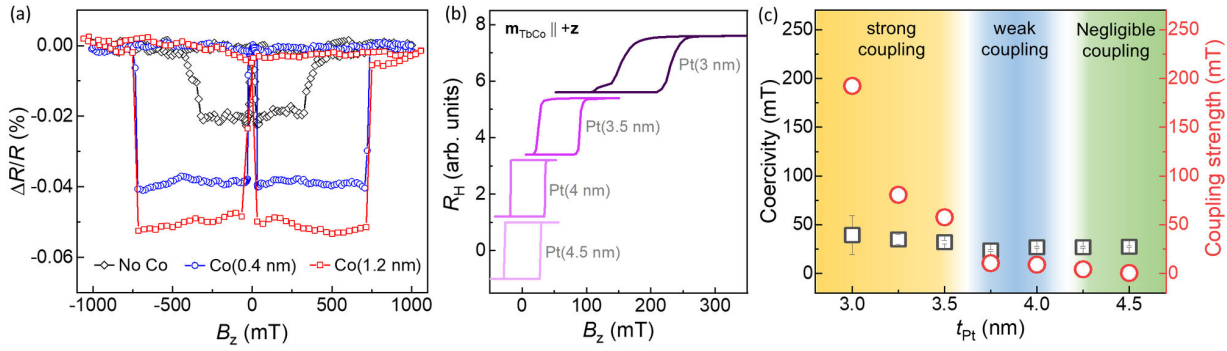

(a) Pt(4)/Co(1.2)/Cu(3.1)/Co(1.2)/TbCo(16 nm)

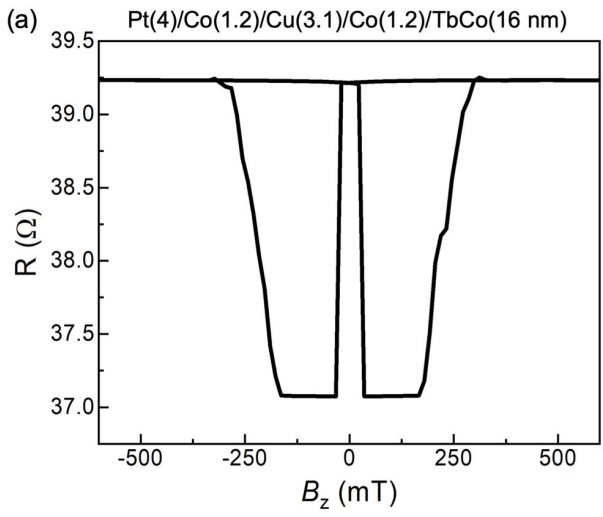

(b)

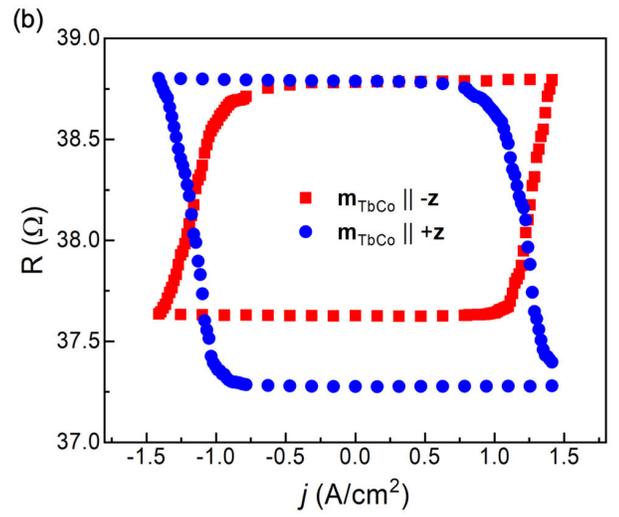